\documentstyle[twocolumn,aps]{revtex}
\newcommand{\be}{\begin{equation}}
\newcommand{\ee}{\end{equation}}
\newcommand{\bea}{\begin{eqnarray}}
\newcommand{\eea}{\end{eqnarray}}
\begin{document}
\draft
\title{\bf Domain Structures in Fourth-Order Phase and\\ Ginzburg-Landau
Equations}
\author{David Raitt$^+$}
\author{Hermann Riecke}
\address{Department of Engineering Sciences and Applied Mathematics\\
Northwestern University \\
Evanston, IL 60208, USA \\}

\maketitle

%\widetext

\begin{abstract}
In pattern-forming systems, competition between patterns with different
wave numbers can lead to domain structures, which consist of regions
with differing wave numbers separated by domain walls.  For domain
structures well above threshold we employ the appropriate phase
equation and obtain detailed qualitative agreement with recent
experiments.  Close to threshold a fourth-order Ginzburg-Landau
equation is used which describes a steady bifurcation in systems with
two competing critical wave numbers. The existence and stability regime
of domain structures is found to be very intricate due to interactions
with other modes.
In contrast to the phase equation the Ginzburg-Landau equation allows a
spatially oscillatory interaction of the domain walls. Thus, close to threshold
domain structures need not undergo the coarsening dynamics found
in the phase equation far above threshold, and can be stable
even without phase conservation. We study their regime of stability
as a function of their (quantized) length. Domain structures are related to
zig-zags in two-dimensional systems. The latter are therefore expected
to be stable only when quenched far enough beyond the zig-zag
instability.

\end{abstract}

\today

Running title: Stability of Domain Structures

Submitted to Physica D.

This paper is also available  from {\it patt-sol@xyz.lanl.gov} using {\it get
9402003}
in the subject field.

\pacs{PACS numbers: 47.20.Ky, 47.20.-k, 05.70.Ln}

\narrowtext
%\newpage
\section{Introduction}

The formation of steady spatial structures in systems far from equilibrium
has been studied in great detail over the past years, the classical
examples being Rayleigh-B\'enard convection and Taylor vortex flow
\cite{m90,ch93}.  In quasi-one-dimensional geometries they usually share a
common feature: the stable structures that arise after the decay of
transients are strictly periodic in space and if they are weakly perturbed
they relax {\it diffusively} back to the periodic state.  This relaxation
has been investigated experimentally in various systems
\cite{cs82,g85,wa91} and the results agree with theoretical results based
on the phase-diffusion equation \cite{rp87,lr90,pr91}.  This equation describes
the
slow dynamics of the local phase, the gradient of which is the local wave
number.  The structures are generally stable over a range of wave numbers
which is limited (at least) by the Eckhaus instability \cite{e65}.  The
Eckhaus instability is
characterized by a vanishing of the phase diffusion coefficient and leads
usually to a new stable periodic state with a different wave number.  This
process involves phase slips, i.e.\ the destruction (or creation) of one or
more roll pairs in convection, say. The Eckhaus instability has been
investigated in great detail \cite{kz85,lg85,kzs88,brs91}, in particular in
Taylor vortex flow \cite{dca86,rp86}.

Recently, it has been pointed out in theoretical work that a vanishing of the
diffusion
coefficient does not necessarily invoke a phase slip
\cite{bd89,r90a,r90b,bd90,dlb90}.  Instead, under certain conditions, the
structure can become inhomogeneous and evolve to a stable domain structure
which exhibits different wave numbers in different parts of the system and
which does {\it not} relax to a periodic structure.  Such a structure is
illustrated in fig.\ref{f:typsol}.  The thick line denotes the local wave
number and the thin line a typical physical quantity such as the mid-plane
temperature in convection.
%This situation has been treated successfully
%using a higher-order phase equation \cite{bd89,r90a,r90b,bd90,dlb90}.
% if the figure shows oscillations the last sentence should be omitted.

Experimentally, inhomogeneous structures have been found in a variety of
systems. Most of them involve at least one time-dependent structure, e.g.\
counterpropagating traveling waves (or spirals) \cite{als86,mhaw90},
(turbulent) twist vortices amidst regular Taylor vortices \cite{ba86},
localized traveling-wave pulses in binary-mixture convection \cite{bks90}
and steady Turing patterns amidst chemical traveling waves \cite{pddkdwdb92}.
Recently, however, domain structures involving only steady convection rolls
of two different sizes have been observed in Rayleigh-B\'enard convection
in a very narrow channel \cite{hvdb92}. It has been suggested that these
structures may be related to the phase-diffusion mechanism discussed above
\cite{bd89}.

Similar phenomena are relevant in two-dimensional systems. There, roll
(or stripe) patterns usually undergo an instability at small wave
numbers in which the phase-diffusion coefficient for perturbations {\em
along} the rolls goes through zero. This leads to an undulatory
deformation of the rolls which can grow into a zig-zag pattern. Such
patterns are related to domain structures with the `zigs' corresponding
to domains of one wave number and the `zags' corresponding to domains
of a  different wave number.

Here we investigate steady domain structures from two points of view.
For structures well above threshold a description of domain structures
is obtained using the appropriate 4th-order phase equation introduced
earlier \cite{bd89,r90a}.  In sec.\ref{s:phase} we perform a detailed
study of this equation. We emphasize the bifurcation that leads to the
domain structure and its connection to the Eckhaus instability. These
results are compared with recent experiments \cite{hvdb92}.

Domain structures can also form close to threshold if the basic state
becomes unstable to two periodic patterns with different wave numbers.
If these two patterns arise at similar values of the control parameter,
the resulting competition can be described by a Ginzburg-Landau
equation.  In sec.\ref{s:ampeq} we introduce and investigate the
appropriate equation.  We present a detailed analysis of the stability
regime of domain structures for periodic boundary conditions. In part
of the parameter regime they can again be described by the above phase
equation.  Surprisingly, we find no trace of the breakdown of the phase
equation in the stability behavior of the domains.  In fact, the
domains can be stable even if the local wave number is well below the
neutral curve, where the phase equation clearly does not apply.

Within the phase equation, domain structures are only stable if the
total phase, i.e.\ the total number of cells, is conserved.  In
sec.\ref{s:subramp} we therefore study the Ginzburg-Landau equation
with a spatially ramped control parameter which allows the total phase
to change.  We find stable domain structures even in this general case
and explore their regime of existence.  In sec.\ref{s:2dapp} we briefly
address the relevance of our results to  two-dimensional patterns and
discuss the zig-zag instability and the stability of the resulting
zig-zag patterns in isotropic systems, e.g.\ Rayleigh-B\'enard
convection and Turing structures \cite{db92}, and in anisotropic
systems like electroconvection in nematic liquid crystals \cite{rj86}.

\section{Domain Structures within the Phase Equation}
\label{s:phase}

In this section we consider steady one-dimensional periodic structures
away from threshold. The slow dynamics of long-wavelength
perturbations of such structures can be described by an equation
for the local phase of the structure. To derive the phase equation one
assumes the structure to be close to periodic and expands the relevant
fields, e.g.\ the fluid velocity, according to
\be
v(x,t) = v_0(\epsilon^{-1} \varphi(X,T)) +
 \epsilon v_1(\epsilon^{-1} \varphi(X,T)) + h.o.t.,
\label{e:vexp}
\ee
where the slow variables $X=\epsilon x$ and $T=\epsilon^2t$, $\epsilon\ll 1$
have been
introduced \cite{m90}. Inserting this expansion into the basic
equations, e.g.\ the Navier-Stokes equation, yields at $O(\epsilon)$
the solvability condition
\be
\partial_T\varphi = {\cal D}(q,R)\partial_X^2 \varphi.
\label{e:phase1}
\ee
Note that the phase equation (\ref{e:phase1}) is nonlinear through the
dependence of the diffusion coefficient on the local wave number $q =
\partial_X \varphi$. In addition, we have indicated the dependence of
${\cal D}$ on a control parameter, such as the Rayleigh number $R$ in
convection. As long as the diffusion coefficient ${\cal D}(q,R)$ is
positive, perturbations decay diffusively and the structure is stable
with regard to long-wavelength perturbations.

If ${\cal D}(q,R)$ becomes negative the structure loses stability and
perturbations in the local wave number grow. Usually this leads to a
phase slip and eventually to
a periodic state with a different wave number. If, however, ${\cal D}(q,R)$
becomes
negative in the center of the band of stable wave numbers, the perturbations
can saturate
and a domain structure arises. This is quite naturally the case when the
neutral curve
has two minima (see secs.III,IV below) \cite{r90a,RiSi94}. It can, however,
even occur if the neutral curve has a single minimum \cite{r90a,r90b}.

 To ensure well-posedness of the phase equation for ${\cal D}(q,R)<0$
it is necessary to retain higher order derivatives. Introducing an
additional slow time $\tau=\epsilon^4 t$ and expanding the phase around a
spatially periodic structure with wave number $q_0$, $\varphi(X,\tau) =
q_0 X + \epsilon\phi(X,\tau)$, one obtains
\be
\partial_\tau\phi = (D + E \partial_X \phi + F (\partial_X \phi)^2)
\partial_X^2 \phi -G \partial_X^4 \phi,
\label{e:phase2}
\ee
where
\bea
\epsilon^2D={\cal D}(q_0,R),\ \ \ \epsilon E=\partial_q{\cal D}(q_0,R),\ \ \\
\nonumber
F=\partial^2_q{\cal D}(q_0,R),\ \ \ G>0.
\eea
In anticipation of the situation relevant to this paper, we assume $D$,
$E$, and $F$ to be ${\cal O}(1)$.  These coefficients can also be
determined in a linear stability analysis of the periodic state
from the behavior of the growth
rate at small values of the Floquet parameter (cf.  (\ref{e:growth})
below).  The wave number $Q=\partial_X \phi\equiv(q-q_0)/\epsilon$
satisfies
\be
\partial_TQ = \partial_X^2(DQ + \frac{E}{2}Q^2 + \frac{F}{3}
Q^3-G\partial_X^2Q).
\label{e:phase3}
\ee
This equation is the Ginzburg-Landau equation for a conserved order
parameter $Q$\cite{ko82}. The static solutions to (\ref{e:phase3}) are
most easily understood by rewriting the right-hand side as the equation
for a particle in a potential,
\bea
\partial_X^2Q&=&-\partial_Q U(Q), \nonumber \\
U(Q)&=&-\frac{D}{2G}Q^2-\frac{E}{6G}Q^3- \frac{F}{12G}Q^4+CQ
\label{e:pot}
\eea
with the integration constant $C$. Usually ${\cal D}(q)$ is concave
downward implying $F<0$. This yields a potential as sketched in
fig.\ref{f:pot}a, which allows two kinds of solutions. The three extrema
correspond to spatially periodic structures, of which only the
maximum yields a stable solution within the full equation (\ref{e:phase2}).
In addition, there exists a family of unstable, oscillating (in $X$)
solutions, which correspond to structures with spatially modulated wave
numbers. It contains, as a limiting case, a homoclinic orbit which
describes a system with a localized `dip' (or `hump') in
the wave number and an associated amplitude depression
\cite{kz85}. By contrast, if $F>0$
(fig.\ref{f:pot}b) two maxima are possible allowing also a heteroclinic
orbit connecting two {\it stable} periodic structures. In this paper we
concentrate on such solutions.

The connection between the different solutions is revealed by a
bifurcation analysis close to the onset of the instability of the
periodic structure.  Since (\ref{e:pot}) can be solved exactly in terms
of elliptic integrals it is possible to perform the expansion in the
exact solution. This approach emphasizes the importance of phase
conservation and is demonstrated in \cite{r94}.

It is easier, however, to expand the phase in (\ref{e:phase2})
\be
\phi=\eta A_1 e^{iPX} + \eta^2(A_2 e^{iPX} + B_2 e^{2iPX} ) + h.o.t. + c.c.
 \label{e:phiexp}
\ee
and derive an evolution equation for $A_1$. Here $P=\frac
{2\pi}{L}\cdot n$, $n$ integer, and the amplitudes depend on the slow
time $\hat{T}=\eta^4 t$, $\eta \ll 1$. No term independent of $X$
arises in (\ref{e:phiexp}) because (\ref{e:phase2}) is invariant under
$\phi \rightarrow \phi+const$.  At fifth order one obtains
\bea
\partial_{\hat T} A_1 &=& -D_4 P^2 A_1 + P^2(\frac{E_2 E_v}{3G}-F_2P^2)
|A_1|^2A_1 \nonumber \\
&-&\frac{5}{216}\frac{E_v^4}{G^3}|A_1|^4A_1,\label{e:fifth}
\eea
where the coefficients
\be
D=D_v+\eta^4D_4, \ \ E=E_v+\eta^2 E_2,\ \ F=F_v+\eta^2 F_2
\label{e:phascrit}
\ee
have been expanded around the tricrital point
\be
D_v=-GP^2,\ \ \  E_v^2=6GF_vP^2. \label{e:vert}
\ee
Thus, the instability, which is signified by a sign change of the
diffusion coefficient, need not always induce a phase slip, as it
usually would in the Eckhaus instability. Instead, if ${\cal D}(q)$
is negative in only a (small) region around a minimum, i.e.\ if $F\equiv
\frac{1}{2} \partial_q^2{\cal D}(q)$ is sufficiently positive, the
bifurcation becomes supercritical and the domain structures, which
correspond to an oscillatory mode of the `particle in the potential',
are stable.  In the subcritical case the domain structures persist up
to the saddle-node given by
\be
D_4 = \frac{6GP^4}{5E_v^4}(E_v E_2 - 3F_2GP^2)^2. \label{e:sn}
\ee

To make contact with experiments \cite{hvdb92}, we model the diffusion
coefficient ${\cal D}(q)$ in the vicinity of a minimum in $q$,
\bea
{\cal D}(R,q)&=&{\cal D}(R_m,q_m)+\partial_R {\cal D}\cdot(R-R_m) \nonumber \\
&+& \frac{1}{2}\partial_q^2 {\cal D}\cdot(q-q_m)^2.
\label{e:dexp}
\eea
At a Rayleigh number $R$, the parameters $D, E$ and $F$, which are
calculated by expanding ${\cal D}(q,R)$ around the periodic solution at
wave number $q_0$, are then given by
\bea
D&=&\partial_R {\cal D}\cdot(R-R_m) + \frac{1}{2}\partial_q^2 {\cal
D}\cdot(q_0-q_m)^2, \nonumber \\
E&=&\partial_q^2 {\cal D}\cdot(q_0-q_m),\ \ \ F=\partial_q^2 {\cal D}.
\label{e:def}
\eea
Here we have chosen $R_m$ such that ${\cal D}(q_m,R_m)=0$.  A typical
phase diagram for such a case is sketched in fig.\ref{f:expresult}a.
The thin line denotes the onset of the instability, whereas the thick
lines give the locus of the saddle nodes.  When raising $R$ the
instability first appears at $q_m$ and is supercritical.  Just before
the transition the phase diffusion coefficient is still positive but it
is very small over some range of wave numbers, thus irregularities in
the structure are expected to decay very slowly.  For larger $R$ and
consequently away from the minimum the bifurcation becomes
subcritical.  The width $\delta q=2P\sqrt{6G/\partial^2_q{\cal D}}$ of
the supercritical regime depends on the length $L\equiv 2\pi/P$ of the
system and vanishes as $L\rightarrow\infty$.  In the subcritical
regime, there exist two different domain structures in the region
between the Eckhaus instability and the saddle-node line, one of which
is unstable.  This is related to the result of Brand and Deissler who
find two different solutions of this type numerically \cite{dlb90}.

In convection experiments in a very narrow channel Hegseth {\em et al.}
have found stable domain structures with different wave numbers
\cite{hvdb92}.  Brand and Deissler \cite{bd89} suggested that they are
related to the solutions discussed above, a position that is supported
by the present results.  When raising the Rayleigh number Hegseth {\em
et al.} find a transition to domains, which can be supercritical or
subcritical, depending on the average wave number $q_m$.  Their results
are shown in fig.\ref{f:expresult}b for comparison with the predictions
by the phase equation. This figure plots the maximum and minimum wave
numbers of the pattern as the Rayleigh number is increased for two patterns
which
differ in their initial, homogeneous wave numbers. If the initial wave
number is close to $q_m$ there is a smooth (supercritical) transition
to a domain structure as indicated by the circles.  On the other hand,
if the initial wave number is far from $q_m$, the periodic pattern
persists to much higher values of the Rayleigh number and then jumps to
the domain structure. The data for this case is shown by the squares
and is indicative of a subcritical bifurcation.

Near $q_m$ and $R_m$, our analysis suggests that phase diffusion should
become very slow.  This agrees with the experimental finding that the
domain structures do not relax to a periodic pattern within the
observational time scale when $R$ is decreased just below $R_m$
\cite{hvdb92,d92}.  From the above discussion the diffusion coefficient
${\cal D}(q)$ is expected to have a bimodal shape in $q$ and a minimum
at ($R_m,q_m$), and to go through zero linearly in $R$ there.  This
could be tested experimentally with methods similar to those used by Wu
and Andereck in Taylor vortex flow \cite{wa91}.

Within the framework of the phase equation the interaction between
domain walls is purely attractive \cite{ko82}. Therefore, in regimes in
which the phase equation applies domain walls
annihilate each other as long as this is compatible with phase
conservation. Thus, if the boundary conditions enforce phase conservation,
the system will eventually reach a state with only a single domain-wall pair.
If the boundary conditions do not conserve phase, e.g.\ in the presence of
subcritical ramping in the control parameter $R$ \cite{kbjbc82,pr91}, domain
structures exist only as transients \cite{bd89,bd90}.
In general, however, the interaction may also be repulsive,
allowing domain-wall pairs to be stable without phase conservation.
Such a situation can arise near threshold. It is examined in section
\ref{s:subramp} using the Ginzburg-Landau equation
which is introduced in the next section.

\section{Domain Structures Within a Fourth-Order Ginzburg-Landau Equation}
\label{s:ampeq}

Close to threshold domain structures can arise if the basic state becomes
unstable with respect to two patterns with different wave numbers, i.e.\ if
the neutral curve for the appearance of a pattern has two minima.  If the
wave numbers are close to each other, such a system can be described by a
single Ginzburg-Landau equation.  It should be emphasized that in addition
the minima must be due to the same mode.  This is the case, for instance,
in parametrically driven standing waves \cite{r90a,RiSi94}.  Such a
Ginzburg-Landau equation is fourth-order in space, reflecting the two
minima of the neutral curve,
\be
\partial_{\tilde{T}} \tilde{A} = \tilde{D}_2 \partial_{\tilde{X}}^2
\tilde{A} +
i\tilde{D}_3\partial_{\tilde{X}}^3 \tilde{A} -
\tilde{D}_4\partial_{\tilde{X}}^4 \tilde{A} + \tilde{\Sigma} \tilde{A}
- \tilde{\Gamma} |\tilde{A}|^2\tilde{A}.\label{e:GL4t}
\ee
The convective amplitude $\tilde{A}$ gives a typical quantity, e.g.\ the
vertical velocity in convection, via
\be
\tilde{v}_{\tilde{z}}(\tilde{x},\tilde{z},\tilde{t})=\delta
e^{i\tilde{q}_c\tilde{x}} \tilde{A}(\tilde{X},\tilde{T}) f(\tilde{z}) +
h.o.t. + c.c.
\label{e:texp}
\ee
where $\tilde{X}=\delta \tilde{x}$ and $\tilde{T}=\delta^4 \tilde{t}$ are
slow variables, $\delta \ll 1$, and $f(\tilde{z})$ is the appropriate
vertical eigenfunction. In order for (\ref{e:GL4t}) to be asymptotically
valid $\tilde{D}_2 \partial_{\tilde{X}}^2 \tilde{A}$ and
$i\tilde{D}_3\partial_{\tilde{X}}^3 \tilde{A}$ have to be of the same order
as $\tilde{D}_4\partial_{\tilde{X}}^4\tilde{A}$. Periodic solutions of this
Ginzburg-Landau equation and their stability have been studied by Proctor
\cite{p91}.

In the following we solve (\ref{e:GL4t}) numerically for a finite system of
length $\tilde{L}$ with periodic boundary conditions.  We are interested in
particular in the behavior of the domains when the wave-number gradients
become steep.  To achieve this we scan across the full instability regime
of the {\it periodic} pattern by changing the average wave number of the
domain structure.  To this end we investigate the dependence of the system
on its length $\tilde{L}$.  To keep $L$, the nondimensionalized system
length, as a parameter we scale lengths using the critical wave number
${\tilde q}_c$ rather than ${\tilde L}$,
\bea
\tilde{X}&=&X/\tilde{q}_c,\ \ \tilde{q}=\tilde{q}_c (1 + \delta Q), \ \
\tilde{L}=L/\tilde{q}_c, \nonumber\\
\tilde{A}&=&A{\tilde{q}_c}^2\sqrt{\tilde{D}_4/\tilde{\Gamma}},\ \
\tilde{T}=T/(\tilde{D}_4{\tilde{q}_c}^4),\nonumber \\
\tilde{D}_2&=&D_2\tilde{D}_4{\tilde{q}_c}^2,\ \
\tilde{D}_3=D_3\tilde{D}_4{\tilde{q}_c},\ \
\tilde{\Sigma}=\Sigma\tilde{D}_4{\tilde{q}_c}^4.
\label{e:scal}
\eea
With this scaling the Ginzburg-Landau equation becomes
\be
\partial_T A = D_2 \partial_X^2 A +
iD_3\partial_X^3 A - \partial_X^4 A + \Sigma A - |A|^2A.\label{e:GL4}
\ee
Either $D_2$ or $\Sigma$ could still be scaled away, leaving the quantity
$D_2^2/\Sigma$ as the primary control parameter.  We have chosen to retain
both parameters to simplify scanning each of them across zero.  As pointed
out by Tuckermann and Barkley, periodic boundary conditions for the
physical quantities like $\tilde{v}_{\tilde{z}}$ imply non-periodic
boundary conditions for the amplitude $A$ \cite{tb91},
\be
A(X+L)e^{iL/\delta}=A(X).\label{e:bc}
\ee
Thus the conserved slow phase is the total slow phase $\Delta\Phi$ of
the pattern, not only that of the slowly varying amplitude and is given
by
\be
\Delta\Phi=L+\delta \int_0^L Q dx. \label{e:tphase}
\ee
For small $\delta$, changing the length has a very strong effect on the
amplitude $A$, since small changes in the physical wave number $\tilde q$
imply large changes in the reduced wave number $Q$.  In particular, with
these boundary conditions the reduced wave number can be scanned across $0$
by changing the length of the system.  With periodic boundary conditions on
$A$, for which $Q=\frac{2\pi}{L}\cdot n$, this would not be possible.

The neutral stability curve and Eckhaus boundary for (\ref{e:GL4}) are
shown in fig.\ref{f:eckns} for $D_3=0$ in the limit of infinite system
length. The Eckhaus boundary is obtained by a linear stability analysis
of the periodic solution $A=A_1 e^{iQX}$ that yields the growth rate
$\sigma$ of sideband disturbances with wavenumber $Q \pm P$,
\bea
%\sigma&=&-P^4-(D_2+6Q^2)P^2-\Sigma+Q^2D_2+Q^4\pm  \label{e:growth}\\
%& &\sqrt{(\Sigma-Q^2D_2-Q^4)^2
%+4P^2Q^2D_2^2+16P^2Q^2(P^2+Q^2)D_2+16P^2Q^2(P^2+Q^2)^2}. \nonumber
\sigma&=&-P^4-(D_2+6Q^2)P^2-\Sigma+Q^2D_2+Q^4\pm  \label{e:growth}\\
& &\sqrt{(\Sigma-Q^2D_2-Q^4)^2
+4P^2Q^2(D_2 + 2 (P^2+Q^2)^2)}. \nonumber
\eea
The Eckhaus curve is obtained from this in the limit $P \rightarrow 0$.
In each of the two minima of the neutral curve the stability of the
periodic pattern is governed by the usual Eckhaus criterion. Note that for
$Q=0$, i.e. at the local maximum of the neutral curve, the periodic pattern is
unstable for {\it all} $\Sigma$. The width of this unstable region is
$\Delta Q = \sqrt{-D_2/6}$. The term ``inner Eckhaus boundary'' used
below will refer to those parts of each Eckhaus boundary for which
$|Q|<|Q_{min}|$, while ``outer Eckhaus boundary'' will refer to those parts
for which $|Q|>|Q_{min}|$, where $Q_{min}=\pm\sqrt{-D_2/2D_4}$ are the
minima of the neutral-stability and Eckhaus curves.

In a finite system there is no longer a continuum of allowed wave numbers,
instead there is a discrete spectrum $\{Q_n\}$ of them. This also implies
that $P$ is quantized in a similar manner. Using the chosen scaling
(\ref{e:scal}), $P$ is constrained to integer multiples of $\frac{2\pi}{L}$. In
this case all allowed periodic solutions can be stable. Eq.(\ref{e:growth})
shows that for this to occur it is necessary that the wave numbers are
spaced widely enough to straddle the unstable region in the center, i.e.
$D_2/P^2 > -5/2$, and sufficient that $Q=0$ itself is stable, which happens
for $D_2/P^2 >-1$. The latter is the case if the adjacent allowed wave numbers
$Q=\pm 2\pi/L$ lie outside the neutral curve. Again, a periodic solution
that is unstable at onset retains its instability for all values of
$\Sigma$.

Keeping $D_2$ fixed, we have as control parameters the total slow phase of
the system $\Delta\Phi$, the system length $L$, and $\Sigma$, which can be
considered a reduced Rayleigh number.  The domains are characterized by the
difference $\Delta Q$ between the largest and the
smallest wave number in the solution which thus serves as an order
parameter (cf.\ fig.\ref{f:typsol}).

The previous work done on the phase equation (cf.  sec.\ref{s:phase})
gives some indication of what to expect in these simulations.  For
large $\Sigma$ and small $D_2$ the phase equation (\ref{e:phase2}), and
therefore its amplitude equation (\ref{e:fifth}), is valid. In that
region of parameter space it is expected that a branch of unstable
domain structures bifurcates from the inner Eckhaus boundary and
undergoes a saddle-node bifurcation to a branch of stable domain
structures (cf. fig.\ref{f:expresult}). On the other hand, for values
of the parameters near the minimum of the neutral-stability curve, the
solutions can be described using the phase equation appropriate to the
typical Eckhaus case. Thus it is expected that there is an unstable
branch that bifurcates subcritically from the inner Eckhaus boundary
and combines with an unstable branch bifurcating subcritically from the
outer Eckhaus boundary. In the following we concentrate on the
transition region between these two types of behavior.

To calculate domain structures and examine their stability we solve the
Ginzburg-Landau equation (\ref{e:GL4}) with boundary conditions
(\ref{e:bc}) numerically.  The solutions are found using Newton's
method and their stability is determined using inverse iteration.  The
average reduced wave number $\bar Q = \int_0^L Q dX /L$ is used as a
control parameter.  To change this it is most straightforward to alter
$L$ by changing the grid spacing $dx$ ($\approx 0.25$ in this
section).  The additional parameters are given by $\Delta\Phi \approx
113.7$, $\delta=0.1$, $D_2=-1.0$ and $\Sigma$ chosen as indicated. We
concentrate on the symmetric case $D_3=0$.

A set of four solutions at representative values of the two control
parameters are shown in fig.\ref{f:foursol}. The average reduced wave
number $\bar Q$ remains constant in the columns while $\Sigma$ remains
constant in each row. The local wave number is marked by the heavy
solid line, whereas $Re(e^{iq_cx}A)$ with $q_c=1$ is denoted by the thin line.
As is
apparent in fig.\ref{f:foursol}a,c the dominant effect of increasing
the average wave number $\bar Q$ is a decrease in the length of the
low-wave number region in the center.  A striking feature, which is
particularly important in the next section is demonstrated in
fig.\ref{f:foursol}a,b; the local wave number is non-monotonic and
exhibits spatial oscillations. These oscillations become increasingly
pronounced with decreasing $\Sigma$. We will comment on the relevance
of this below.

A sequence of bifurcation diagrams displaying the transition from the
large-$\Sigma$ to the small-$\Sigma$ behavior is given in
fig.\ref{f:bifdia}.  In fig.\ref{f:bifdia}a $\Sigma=1.2$.  The branch
of the stable domain structure is indicated by the bold line
terminating in the saddle-node bifurcation {\it A} at ${\bar Q} \approx
0.68$.  There it collides with an unstable branch of solutions that
bifurcates off the periodic solution ($\Delta Q=0$) at the inner
Eckhaus boundary (cf.  fig.\ref{f:eckns} at ${\bar Q}\approx 0.43$).
The dotted line represents another branch of unstable solutions that
arises from the outer Eckhaus boundary (${\bar Q}\approx 1.12$)and
undergoes a phase slip as ${\bar Q} \rightarrow 0$ which changes
$\Delta \Phi$ by $2 \pi \cdot \delta$.

Lowering the control parameter $\Sigma$ to $1.0$, introduces new
features (fig.\ref{f:bifdia}b).  The original saddle-node bifurcation
{\it A} remains, but two new saddle-node bifurcations, {\it B} and {\it
C}, have appeared in the stable branch.  These new saddle nodes
mark the ends of a loop that grows as $\Sigma$ is reduced.  In going
through the loop, the wave number changes from having a local maximum
and two local minima in the low-wave-number domain (as in
fig.\ref{f:foursol}b) to having only a single minimum in that domain.
Additionally, the branch from the outer Eckhaus boundary moves closer
to the stable solution branch.

Reducing the parameter to $\Sigma = 0.9$ (fig.\ref{f:bifdia}c) produces
significant changes in the bifurcation diagram. The outer Eckhaus branch
merges with the branch of stable domain solutions to produce two new
saddle nodes, {\it D} and {\it E}. As $\Sigma$ is lowered further, the
saddle-node pairs {\it C} and {\it D} as well as {\it A} and {\it E} move
together (fig.\ref{f:bifdia}d) and annihilate leaving only one surviving
saddle node, {\it B}.

Lowering $\Sigma$ to $0.3$ produces no qualitative change in the behavior
of the system.  There continues to be only a single saddle node, at the tip
of the loop (fig.\ref{f:bifdia}e).  An additional (dashed) line is
indicated in this figure.  It represents another solution branch that
undergoes a phase slip of $4\pi\cdot\delta$ as $\bar{Q}
\rightarrow 0$. Comparing this analysis with that of the phase equation
(\ref{e:phase2}) it
is tempting to associate this branch with a phase perturbation $\phi$
of wave number $P=\frac{2\pi}{L}\cdot 2$ (cf.  (\ref{e:phiexp})). It is
therefore expected to merge with the periodic solution for large $\bar
Q$. We were, however, unable to continue it that far, since additional
saddle-node bifurcations arise. In the wave-number region of interest,
it merges with the branch of stable domain structures between $\Sigma =
0.3$ and $\Sigma = 0.25$ producing two new saddle-node bifurcations,
{\it F} and {\it G}, shown in fig.\ref{f:bifdia}f. The latter
annihilates with saddle node {\it B} around $\Sigma=0.2$.  Upon
lowering $\Sigma$ further, additional saddle nodes form and annihilate
pairwise. This is presumably due to the interaction with additional
branches corresponding to phase slips $2\pi\cdot\delta \cdot n, n>2$.

The above numerical results yield the phase diagram shown in
fig.\ref{f:phsdia}. This picture shows one half of the parameter
space and is symmetrically continued for $Q<0$.
Stable domain structures exist in the region above
and to the left of the line formed by the solid circles. Each of them
indicates a saddle-node bifurcation of the stable branch. Above $\Sigma
\approx 1.2$ there exists only a single line of saddle-node
bifurcations {\it(A)}. The first line of new dots ({\it B} and {\it
C}), to the left of line {\it A} denotes the loop first shown in
fig.\ref{f:bifdia}b. The two saddle nodes in the loop are so close
together that they cannot be resolved in this figure. For smaller
$\Sigma$ ($\approx 1.0$), a hump arises from the creation of saddle
nodes {\it D} and {\it E}, at the merging of the stable branch with the
$2 \pi$-phase-slip branch. These saddle nodes annihilate pairwise,
leaving only one ({\it B}). At $\Sigma \approx 0.3$ another hump
appears, caused by the merging of the $4 \pi$-phase-slip branch and the
branch of stable domain structures, creating saddle nodes {\it F} and
{\it G}. Again, two of the saddle nodes merge and annihilate leaving
only one. Similar humps that appear in the saddle-node line suggest
that this process continues as $\Sigma$ is reduced and $\bar Q$
approaches zero.

The phase equation breaks down when the local wave number approaches the
neutral curve.  As fig.\ref{f:phsdia} shows we have been able to find
stable domain structures for values of $\Sigma > -0.177$.  In this regime,
the wave number is locally below the neutral curve.  Thus the domain
structures exist and are stable well beyond the validity of the phase
equation.  Strikingly, this leaves no trace in the stability properties of
the domain structure.  It is, however, responsible for the oscillatory
behavior of the wave number (cf.\ fig.\ref{f:foursol}).
 It is not clear whether or not
the domain structures exist all the way to the minimum of the neutral curve,
occurring at
$\Sigma=-0.25$.

For simplicity we have concentrated on the special case $D_3=0$
\cite{fn1}.  If $D_3\ne 0$ then one of the two minima in the neutral
curve is lower than the other.  Thus, if the control parameter is raised
adiabatically, domain structures will not arise spontaneously from the basic
state as the critical parameter value is crossed.
Instead a periodic pattern with a wave number corresponding to the lower
minimum will arise.  If, however, the control parameter is quenched into
the region in which there are two separate stable regions, then domain
structures are expected to form.  As shown by Proctor \cite{p91}, the
region of stable wave numbers that arises at the higher value of the
control parameter is closed at the top rather than open as is the case
here.  We do not expect this to have a strong effect, however, in the
regions of parameter space under investigation here.

\section{Domain Structures Without Phase Conservation}
\label{s:subramp}

Within the phase equation the dynamics of domain structures depends sensitively
on the boundary conditions. In particular, the stability of a domain structure
consisting of a single domain-wall pair requires the boundary conditions to
conserve the phase.  It is therefore of interest to study which properties of
such
a domain structure are independent of the boundary conditions within the
Ginzburg-Landau equation. This information will also give insight into the
properties of domain structures consisting of many domain walls. In such
structures adjacent domain-wall pairs can annihilate even if the total
phase is conserved since nearby domains can grow or shrink to compensate for
the loss of phase.

To investigate the behavior of domains in such a situation we use the
fourth-order Ginzburg-Landau equation introduced in the previous
section. Boundary conditions which do not conserve the phase can be
obtained by introducing a smooth subcritical ramp in the control
parameter $\Sigma$ at each end of the system \cite{kbjbc82,pr91}.
Since the amplitude at each end of the system goes then to zero, phase
can be added or removed through the boundaries. Effectively, a ramp
poses boundary conditions on the wave number \cite{kbjbc82,pr91}. In this
system it
selects either of the two minima of the neutral curve $\pm Q_{min}$.
Since these wave numbers are consistent with the wave numbers found inside
each domain, the ramps mimic the effect of distant domain structures.

To simulate the time evolution numerically we use a Crank-Nicholson
scheme with $dx\approx0.034$ and $nx=3200$. The weakness of the
attraction between domain walls suggested such a small grid spacing to
eliminate pinning by the lattice points.  The initial conditions for
all of the simulations performed in this section consist of a step
function with three domains, two of high wave number surrounding one of
low wave number.  The initial conditions are varied by changing the
width of the central (low wave number) domain.

In order to confirm that this ramping is sufficient to eliminate phase
conservation, we start by investigating large values of the control
parameter, $\Sigma=100$ (with $D_2=-1.0$).  In this region of parameter
space, the phase equation governs the dynamics of the pattern and the
domain walls will have purely attractive interaction.  As expected, all of
the different initial conditions tested converged to a periodic pattern
with wave number $Q_{min}$.  This shows in addition that the numerical
parameters selected are sufficient to eliminate pinning on the grid in this
regime.

%beginning of next paragraph to be rewritten:

%The next area of investigation is in the transition region of parameter
%space, between the regions where the phase equation introduced earlier
%\ref{e:phase2} is valid and in the wells of the neutral-stability curve,
%where the well-known phase equation is valid.

We next consider the behavior of domains for lower values of the control
parameter $\Sigma$.  These simulations were performed for $\Sigma=1.0$ and
$D_2=-1.0$.  For these values of the control parameter, close to the
neutral-stability curve, the dynamics is no longer expected to be governed
by the phase equation.  The behavior is indeed quite different from that
observed
for high $\Sigma$.  Here the system evolves to a discrete number of
different states, as shown in fig.\ref{f:S1phase}, where the total phase in
the system (between the ramps) is plotted versus time.  As the initial width of
the
central (low wave number) domain is increased, the total phase of the
initial condition is decreased, leading to the different lines shown in
this plot.  The upper lines are those with only an initially narrow region
of low wave numbers.  They converge to a final state with a single minimum
in wave number.  The next set of lines have a slightly wider region of low
wave number and converge to a final state with two minima in the wave
number.  Likewise the lowest set of lines converge to a state with three
minima in the wave number.  The corresponding final states are shown in
fig.\ref{f:3states}.  These results show that the width of the domain
structures is quantized and the final states can be characterized by the
number of minima in the wave number.

We now turn to the transition region between the low-$\Sigma$ regime,
where domain structures exist, and the large-$\Sigma$ regime, where
they do not. In the next section, we will relate the stability regime
of domain structures to that of zig-zag patterns in two-dimensional
systems.  For that application it is more natural to change
$\left|D_2\right|$ than $\Sigma$ since the former is directly related
to the roll-spacing of an initial straight-roll pattern.  These two
control parameters are equivalent since (\ref{e:GL4}) has only a single
control parameter, the ratio $D_2^2/\Sigma$. We therefore study the
dependence on $D_2$ instead of on $\Sigma$.

In order to determine the value of the control parameter at which the
various states disappear, we create initial conditions consisting of
domain structures with different widths for $D_2=-1.0$ and
$\Sigma=1.0$.  Then $D_2$ is increased adiabatically and a new steady
solution is determined using Newton's method.  The values of $D_2$ and
the total phase $\Delta\Phi$ are recorded for each step.  Figure
\ref{f:stability} summarizes these results, plotting $\Delta \Phi$
against $D_2$.  As $D_2$ is increased, the total phase in each state
decreases.  This is mainly because, the wave number in each of the
domains, $Q_{min}$, decreases in absolute value as $\left|D_2\right|$
decreases while the length of the domains stays essentially fixed.
Eventually, each domain structure vanishes through a saddle-node
bifurcation.  These bifurcations occur at different values of the
control parameter for different size domains.  The value of $D_2$ at
which the bifurcation occurs increases for the domains with one to four
minima and then decreases again for the state with five minima.  The
investigation of larger domains was not feasible with our finite
difference code.

In the region of parameter space where the domain structures are stable,
the interaction between the domain walls must change sign as a function of
their distance.  In principle, at large distances this interaction can be
described using perturbation methods on heteroclinic orbits
\cite{cer87,ems90} and in simpler systems it has been shown that a
spatially oscillatory behavior of the domain wall can lead to a locking of
adjacent walls \cite{cer87}. The asymptotic behavior of the solution far away
from
the domain wall can be determined through a linearization around the
corresponding
periodic solution of (\ref{e:GL4}). For the periodic
solution with wave number $Q_{min}$ such an analysis reveals
spatially oscillatory eigenmodes, but also modes that decay
monotonically in space \cite{r94}. Correspondingly,
in the numerical simulations the wave number
is clearly a combination of oscillatory and monotonic decaying functions for
small $\left|D_2\right|$. Thus, for adjacent domain walls to lock into each
other
the existence of an oscillatory mode is not sufficient; it has to dominate the
monotonic
mode.

At $D_2 \approx -0.35$ the spatial decay rates of the two modes are the same
(for $\Sigma=1.0$). For larger $D_2$, the decay rate of the
monotonic eigenmode is lower and it will dominate the oscillatory mode
so the dynamics will be that predicted by the phase equation. For
$D_2<-0.35$ the oscillatory mode has a smaller decay rate than the
monotonic mode so repulsive as well as attractive interaction is
possible.  The oscillations are clearly illustrated in
fig.\ref{f:foursol}a,c where it is apparent that raising $\Sigma$
(lowering $D_2$) makes the oscillations near the domain wall much
stronger. This argument suggests that domain structures could exist up to
$D_2\approx-0.35$ and that wide ones persist to smaller
values of $|D_2|$ than narrow ones.  This would be in agreement with the
results for the states with one to four minima, but it does not explain
the behavior of the five-minima state. Thus, a complete understanding of the
numerical
result seems to require a detailed calculation of the interaction.
%It should be noted, however,
%that this analysis is for very long domain states in which the
%exponential tail is the dominant feature. The shorter states  discussed
%here may not satisfy this condition.

%These results suggest that if one were to start with a random array of
%kinks in a suitably long system and were to increase $D_2$ the shorter
%states would disappear in order. Eventually one would end up with either
%a state that had only a single domain-wall pair (in the case when there
%is phase conservation) or a periodic pattern.

The locking described in this section does not occur within the
fourth-order phase equation (\ref{e:phase2}) despite the presence of the
fourth derivative.  As eq.(\ref{e:phase3}) shows the phase equation is
effectively only second order when considering steady solutions.  If a
sixth-order term $H\partial_X^6Q$ were kept in the phase equation,
oscillatory behavior of the wave number could   be described by the
phase equation for sufficiently large $H$.  For this situation to be
asymptotically valid both $D$ and $G$ have to be vanishingly small at the
same time, which is, however, not to be expected in a typical physical
system.

\section{Domain Structures in Two-Dimensional Systems: Zig-zags}
\label{s:2dapp}

The results discussed in this paper can be extended to describe certain
aspects of two-dimensional patterns.  A common instability of
two-dimensional roll patterns is the zig-zag instability.  It leads to an
undulatory deformation of the pattern along the rolls.  During the
nonlinear evolution the undulations become steeper and domains of parallel
rolls arise. They are separated by domain walls within which the orientation of
the rolls
changes rapidly. Thus, they can be characterized by a wave
vector $(q_x,q_y)$ with $q_y$ positive and relatively constant in a
`zig'-region and negative in a `zag'-region. The behavior of $q_y$
is therefore very similar to that of $Q$ in the domain structures.
Single domain walls between zigs and zags have been
studied previously \cite{MaNe90a}. Here we concentrate on {\it pairs} of domain
walls,
which form a domain structure, and the interaction between them.

In isotropic systems, such as Rayleigh-B\'enard convection, the zig-zag
instability arises generically when the wave number becomes too small,
since the zig-zag pattern leads to an increase in the local wave
number.  While these patterns have not been observed to saturate in
Rayleigh-B\'enard convection \cite{BuAu94}, numerical simulations
\cite{db92} suggest that they can be stable in chemical systems
\cite{cdbd90,os91}.

Zig-zag patterns are also of great importance in anisotropic systems
such as electro-hydrodynamic convection in nematic liquid crystals.
There the long molecules exhibit a preferred direction and thus define
an axis of anisotropy of the fluid.  In these systems there are two
possible primary bifurcations to roll-like structures.  The first is to
a parallel-roll pattern, in which the rolls are oriented normal to the
axis of anisotropy and have wave vector $(q_x,0)$.  Such a pattern is
therefore commonly referred to as a ``normal-roll'' pattern.  The
second type of primary bifurcation is to rolls oriented at an angle to
the director.  Due to reflection symmetry, such rolls can be oriented
in two directions with wave number $(q_x,\pm q_y)$.  This allows the
formation of zig-zag patterns \cite{rj86}.

In a simplifying approach, it is possible to describe zig-zag-patterns
using the phase equation (\ref{e:phase3}) with $Q$
corresponding to $q_y$ \cite{bkkpwz91,kbpz87,b89}.  Thus, only variations in
$q_y$ are considered, while $q_x$ is taken to be constant.  As discussed
above, such a description contains in general no stable zig-zag structures;
instead
there is a coarsening of the pattern until there are no domain walls left
and the pattern consists of only one type of rotated rolls.  Only if
the boundary conditions enforce phase conservation, there will be a single
domain wall left (a pair of domain walls for periodic boundary conditions)
which corresponds to a `zig-zag'-structure (`zig-zag-zig'-structure for
periodic boundary conditions).

In two-dimensional experimental systems the total phase is usually not
conserved.  It is therefore again of interest to investigate the stability
of zig-zags in the absence of phase conservation.
Thus, we consider the appropriate Ginzburg-Landau equation.  For isotropic
systems (as in Rayleigh-B\'enard convection) it is given by \cite{nw69},
\be
\partial_TA=-(i\partial_X+\partial_Y^2)^2A+\lambda A-|A|^2A,
\label{e:iso2d}
\ee
while for anisotropic systems (as in convection in nematics) it reads
\cite{pk86},
\be
\partial_TA=(\partial_X^2-iZ\partial_X\partial_Y^2+W\partial_Y^2-
\partial_Y^4)A+\lambda A-|A|^2A.
\label{e:aniso2d}
\ee
In the anisotropic case, the Lifshitz point, at which the pattern at onset
changes
from normal to oblique rolls, is given by $W=-ZQ_x$.

Concentrating on solutions which are strictly periodic in $X$,
$A=A_1(Y,T)e^{iQ_xX}$, one obtains eq.(\ref{e:GL4}) with
\bea
D_2=2Q_x, \ \ \Sigma=\lambda - Q_x^2 \ \ \mbox{for (\ref{e:iso2d})\ }\\
D_2=W+ZQ_x, \ \ \Sigma=\lambda - Q_x^2 \ \ \mbox{for (\ref{e:aniso2d}).}
\label{e:param2d}
\eea
Note that $Y$ and $Q_y$ in the two-dimensional systems correspond to $X$
and $Q$ in eq.(\ref{e:GL4}) and that now periodic boundary conditions for the
physical quantities imply periodic boundary condtions for $A_1$.
Thus, the domain structures discussed in
sec.III are indeed one-dimensional analogs of zig-zag patterns.

Zig-zag patterns have been studied in some detail for anisotropic systems
\cite{pk86,bkkpwz91,kbpz87,b89}.  These studies found that there is no
continuous transition from normal rolls to zig-zags of small amplitude.
For sufficiently large amplitude, however, stable zig-zags are found to
exist in islands within parameter space.  Within these islands, the zigs
and zags always orient themselves in the direction of maximal growth rate.
Presumably these islands are related to the discrete set of domains with
different widths discussed above.

Our results suggest a mechanism to explain the existence of these islands
of stability.  As eqs.(\ref{e:iso2d}-\ref{e:param2d}) show, these results
apply to both the anisotropic and the isotropic case.  The implications of
the results shown in fig.\ref{f:stability} are presented in
figs.\ref{f:2dstability}a-c.  There the loci of the saddle-node bifurcations of
the different domain states are given in the $\lambda-Q_x$-plane (medium-weight
lines).
These are the lines with $D_2^2/\Sigma=constant$, where the appropriate
constant
for each domain structure is obtained from fig.\ref{f:stability}.
In addition the Eckhaus curve and the onset of the zig-zag instability are
pictured.  The figures differ in the values of $Z$ and $W$, with
fig.\ref{f:2dstability}(b) corresponding to the isotropic case. In each figure,
the behavior of the patterns is similar.
Between the Eckhaus instability (thin dotted line) and the
zig-zag instability (thin dashed line) straight-roll patterns are stable.  As
the zig-zag instability is crossed they become unstable to undulations.
Close to the onset of the instability  the coarsening dynamics
predicted by the phase equation will be observed.  This results in a
pattern that will eventually consist of oblique rolls in only one
direction.  Their wave number will be close to that with maximal growth
rate and they will be stable.

If, however, the pattern is quenched deeper into the zig-zag-unstable
regime the resulting zig-zag pattern will be able to persist, as the
oscillations in $Q_x$ can lock into one another.  The analysis of the
spatial decay rates suggests that it is necessary to quench at least beyond
the thickest solid line, which corresponds to $D_2<-0.35$. The numerical
results indicate
one has to quench all the way beyond the medium-weight solid line, which
corresponds
to the four-minima state to find stable zig-zags.
The different medium-weight lines  denote the stability limits
of `zig-zag-zig' structures with different lengths of the `zag'-domain.
Note that these lines merge with the line denoting the zig-zag-instability
on the neutral curve (thin solid line). Except for $W=0$
the merging occurs therefore
at wave numbers for which the straight rolls are unstable
with respect to the Eckhaus instability. Consequently,
for small $\lambda$
stable, locked domain structures arise only at wave numbers $Q_x$
for which the straight rolls are unstable. Although the Eckhaus-instability
of straight rolls does not necessarily imply that of the zig-zag-structures,
it points to the importance of instabilities which involve
the $X$- as well as the $Y$-dependence and suggests that these may be of
particular
importance for small $\lambda$. In our analysis there
is no allowance for such two-dimensional instabilities. For some
of the domain structures they have been investigated in the anisotropic
case \cite{pk86,bkkpwz91,kbpz87,b89}. In these studies stable zig-zag
structures have been found in certain parameter regimes.

A typical `zig-zag-zig' solution is illustrated in fig.\ref{f:2dpic}a for
$D_2=-1$ and $\Sigma=0$.  It takes the one-dimensional pattern $A(X)$ shown
in fig.\ref{f:2dpic}b and extends it into two dimensions by plotting
$Re(A(X)e^{iQ_yY})$ with constant $Q_y=0.7$.  The lines represent the
zeroes of the pattern, while the dark spots are maxima and minima.  This
pattern is stable.  A signature of the oscillations in $Q_x$ is the
appearance of undulatory deformations in the zig-zag structure near the
domain walls.

% multiple domain structure: $D_2 = -0.72$ and $\Sigma =0.51$ $Q_y=2.8$.
%Only when $D_2$ is
%raised will the narrowest states cease to exist and disappear, leaving a
%coarsened structure.   They become more pronounced when the value of $\Sigma$
%%is
%reduced as is shown clearly in fig.\ref{f:2dpic}c,d for a pair of domain
%walls.

\section{Conclusion}

We have considered two approaches to investigating the existence and
stability of domain structures in pattern-forming systems: a phase
equation and a Ginzburg-Landau equation. Within the phase equation the
structures typically arise through a subcritical bifurcation.  This
approach yields qualitative agreement with various experimental
results in slot convection \cite{hvdb92}. As a conclusive test we suggest
a measurement of the wave-number dependence of the phase diffusion
coefficient ${\cal D}(q)$, which is expected to have a bimodal shape.
The phase-equation approach becomes invalid when the transition region
between adjacent domains becomes too narrow, i.e. when the gradient in
wave number becomes too steep. This is in particular the case near onset.

In the second approach we used a Ginzburg-Landau equation to describe the
competition between patterns with different wave numbers close to
onset.  Such a competition naturally arises in a system with a
double-welled neutral-stability curve, as is, for instance, possible in
parametrically driven standing waves \cite{r90a,r90b,RiSi94}.  This approach
allowed the investigation of domain structures and their stability even for
large wave-number gradients.
% The phrase `instability to phase slip' is not so good as we discussed before:
% from the instability (linear that is) itself one cannot tell whether a phase
% slip will occur or not.
We found domain structures
are stable over their entire region of existence.  Strikingly this region
is very intricate due to the interaction with unstable structures possibly
arising from the outer Eckhaus boundary.

The domain structures exist well beyond the regime of validity of the
phase equation.  A characteristic feature of those which are not
described by the phase equation is the oscillatory behavior of their wave
number.  This allows adjacent domain walls to lock into each other, forming
domain structures which are stable even without phase conservation. We have
numerically
studied the dependence of their regime of stability as a function of the length
of the
domains.

Zig-zag patterns can be considered as two-dimensional analogs of domain
structures.  Therefore, in systems in which the phase is not conserved,
they are not stable at onset of the zig-zag instability. Only further
in the zig-zag-unstable regime can the locking prevent the pairwise
annihilation
of zigs and zags.

This work has been supported by grants from NSF/AFOSR (DMS-9020289,
DMS-9304397) and
DOE (DE-FG02-92ER14303). H.R. acknowledges interesting discussions with
W. Pesch, L. Kramer, and E. Bodenschatz.

$^+$ present address: SCRI, Florida State University B-186,
400 Dirac Science Center Library, Tallahassee, FL 32306.

%\typeout{is your new address correct??????}

\newpage
%\centerline{ FIGURE CAPTIONS}

\begin{figure}
\caption{Typical domain structure solution. The thick line gives
the local wave number, while the thin line represents a typical physical
quantity like the
mid-plane temperature.
\protect{\label{f:typsol}}
}
\end{figure}

\begin{figure}
\caption{Potentials U(q) for (a) $F<0$ (${\cal D}(q)$ concave down)
(b) $F>0$ (${\cal D}(q)$ concave up).
\protect{\label{f:pot}}
}
\end{figure}

\begin{figure}
\caption{Comparison of (a) phase diagram obtained from the phase equation
(\protect{\ref{e:fifth}}) with (b) experimental results obtained by Hegseth
{\em et al} \protect{\cite{hvdb92}}.
\protect{\label{f:expresult}}
}
\end{figure}

\begin{figure}
\caption{Neutral stability and Eckhaus curves for $D_2=-1$ and $D_4=1$.
\protect{\label{f:eckns}}
}
\end{figure}

\begin{figure}
\caption{Four solutions obtained at different values of the
control parameters $\bar{Q}$ and $\Sigma$. The local wave number is marked by
the heavy
solid line, whereas $Re(e^{iq_cx}A)$ with $q_c=1$ is denoted by the thin line.
The order parameter $\Delta Q$ is also indicated.}
\protect{\label{f:foursol}}
\end{figure}

\begin{figure}
\caption{Bifurcation diagrams for various values of $\Sigma$:  a)
$\Sigma=1.2$,
b) $\Sigma=1.0$, c) $\Sigma=0.9$, d) $\Sigma=0.8$, e) $\Sigma=0.3$, f)
$\Sigma=0.25$.
The branches of {\it stable} domain structures are denoted by thick lines.
For details see text.
\protect{\label{f:bifdia}}
}
\end{figure}

\begin{figure}
\caption{Phase diagram: domain structures exist and are stable above and to
the left
of the saddle-node line denoted by solid circles.  The letters denote the
loci of
the respective saddle-node bifurcations shown in
fig.\protect{\ref{f:bifdia}}.  The
dashed and thin lines give the neutral and the Eckhaus curves,
respectively.
\protect{\label{f:phsdia}}
}
\end{figure}

\begin{figure}
\caption{Temporal evolution of the total phase for different initial
conditions in the absence of phase conservation.  The corresponding final
states are shown in fig.\protect{\ref{f:3states}}.
\protect{\label{f:S1phase}}
}
\end{figure}

\begin{figure}
\caption{Coexisting stable final states obtained from the time evolution
shown in fig.\protect{\ref{f:S1phase}}.  The thick line gives the local
wave number in the bulk.  Note that the amplitude
goes to zero toward the boundaries due to the subcritical ramping.
\protect{\label{f:3states}}
}
\end{figure}

\begin{figure}
\caption{Regime of existence of bound pairs of domains walls for
$\Sigma=1$.  The calculation was performed with up to 25,600 points and an
appropriate grid spacing down to $dx\approx0.016$.
The ramped part is up to 3,200 points wide on each side.
\protect{\label{f:stability}}
}
\end{figure}

\begin{figure}
\caption{Extension of one-dimensional results of
fig.\protect{\ref{f:stability}}
to zig-zag patterns for different values of $W$ and $Z$ in
eq.(\protect{\ref{e:aniso2d}}), (a) $W=-0.25$, $Z=1.5$, (b) $W=0.0$, $Z=2.0$
(isotropic case), and (c) $W=0.25$, $Z=1.0$.
To the left of the medium-weight lines zig-zags of different widths are locked:
four-minima state (solid), one-minimum state (dotted).
At the thickest line the monotonic and the oscillatory mode
have the same decay rate. Between it and the zig-zag
instability (thin dashed line), there are no stable straight-roll or zig-zag
solutions within this framework. Thin
lines: neutral-stability curve (solid),
Eckhaus instability (dotted).
Between the zig-zag instability and the Eckhaus instability straight-roll
patterns are stable. \protect{\label{f:2dstability}}
}
\end{figure}

\begin{figure}
\caption{Two-dimensional analog of domain structure: zig-zag pattern (a)
and the domain structure (b) from which it was generated.  Both the wave
number (thick line) and $Re(Ae^{iq_cx})$ with $q_c=1$ (thin line) are shown in
(b).  The
parameter values are $\Sigma =0$, $D_2 =-1$, $Q_y=0.7$.
\protect{\label{f:2dpic}}
}
\end{figure}


\begin{thebibliography}{99}
\bibitem{fn1}This is obtained in the one-dimensional
description of two-dimensional systems, where the odd derivative vanishes
due to reflection symmetry (see (\ref{e:iso2d}) and (\ref{e:aniso2d}) below).
\end{thebibliography}
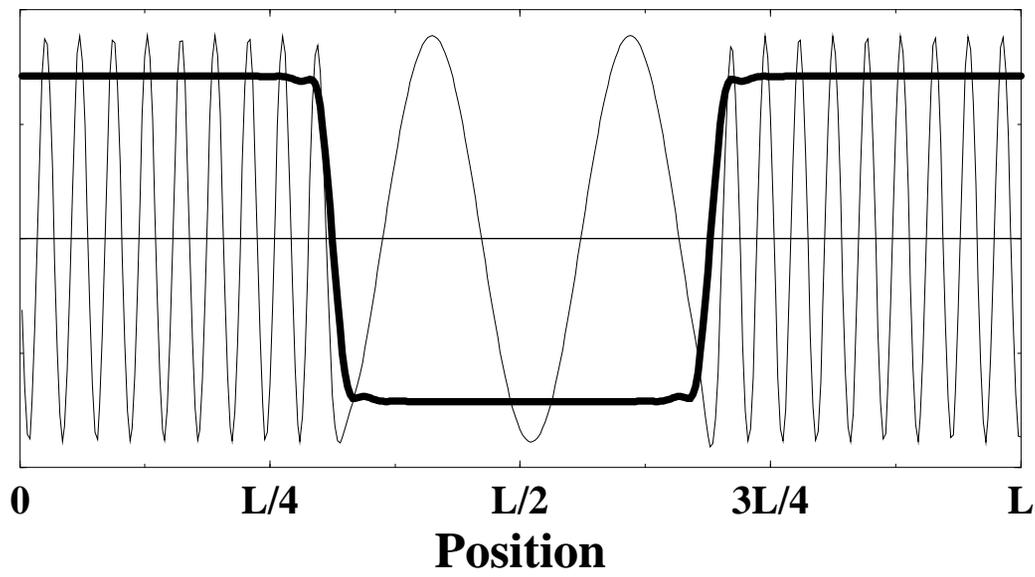
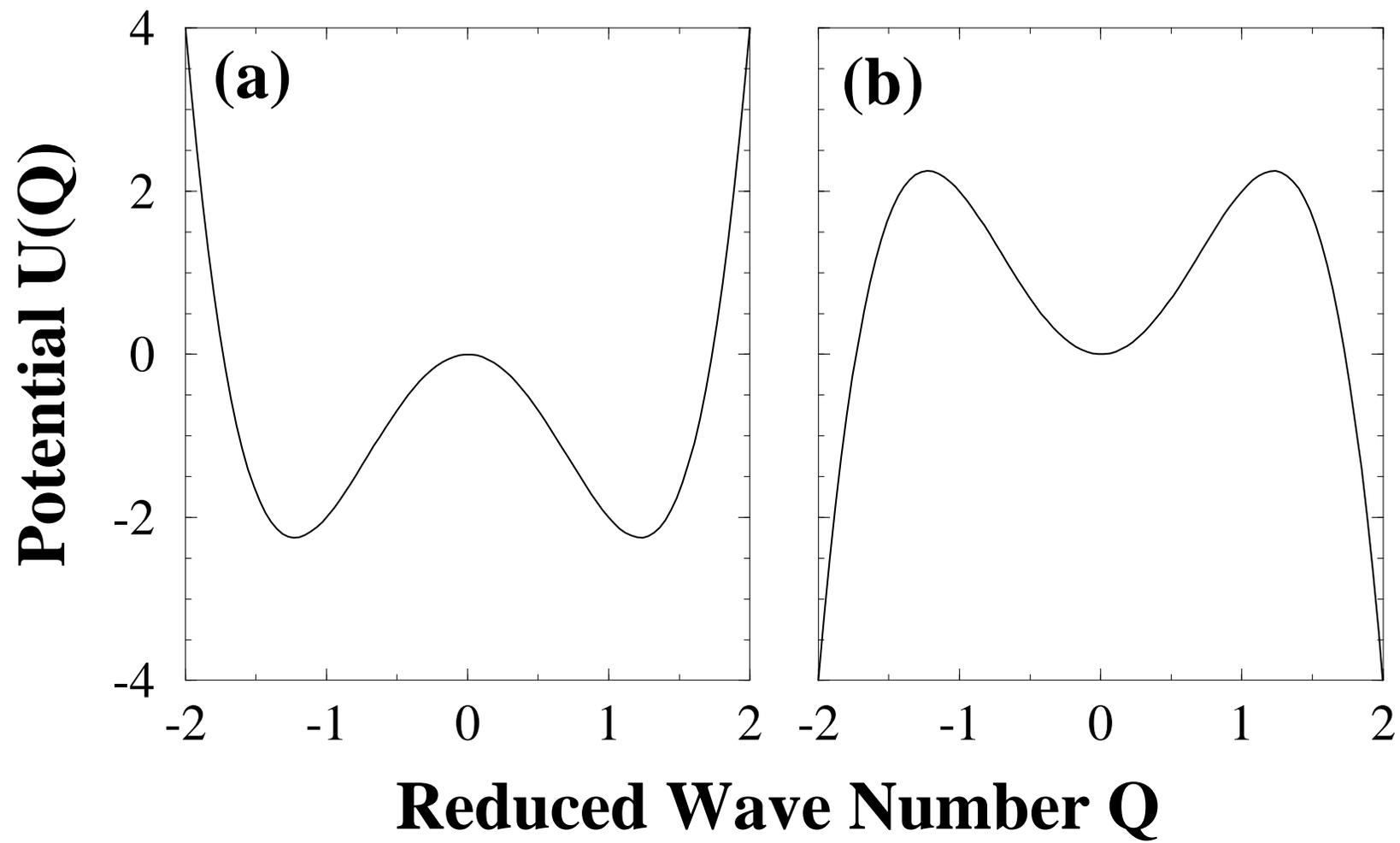
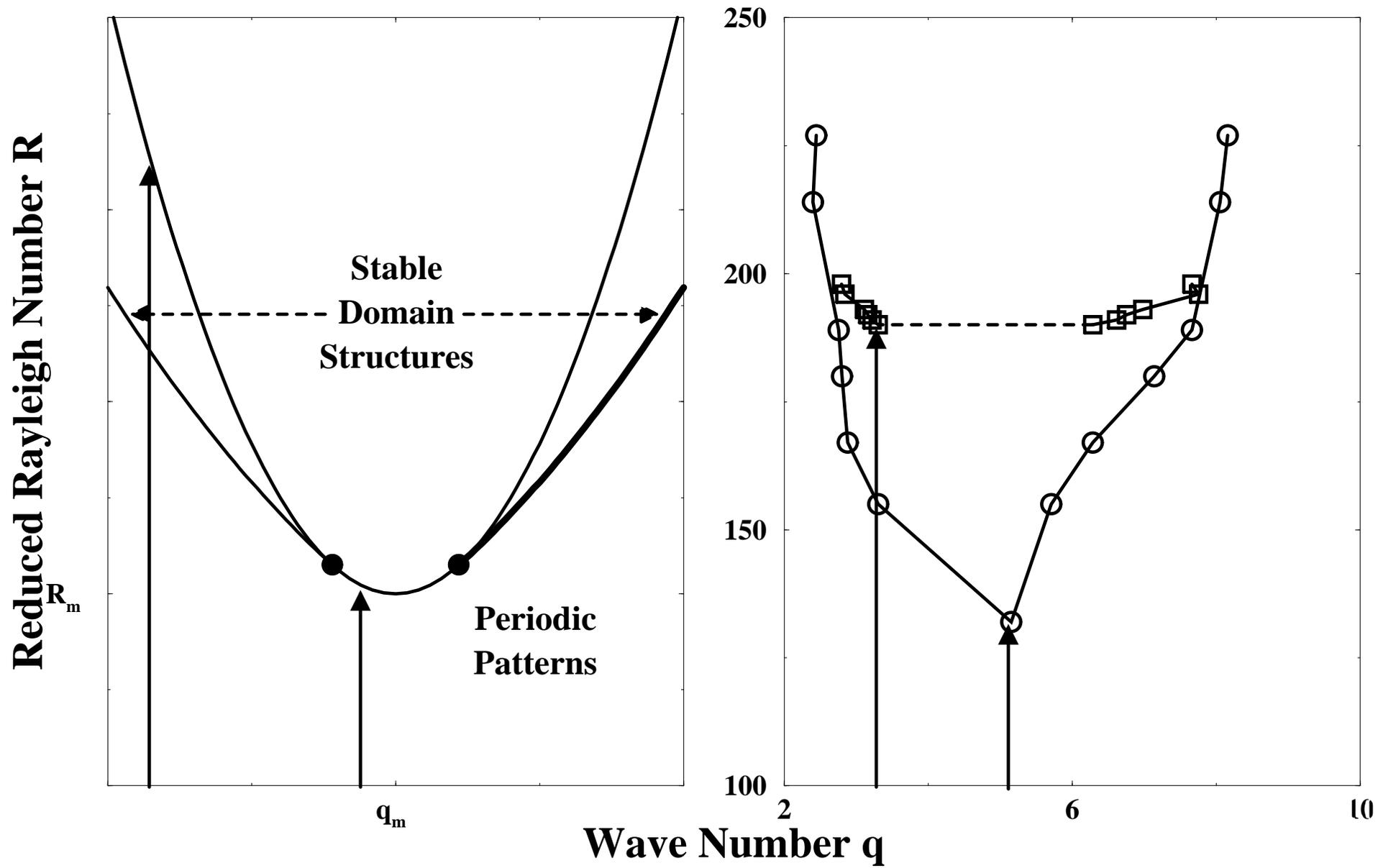
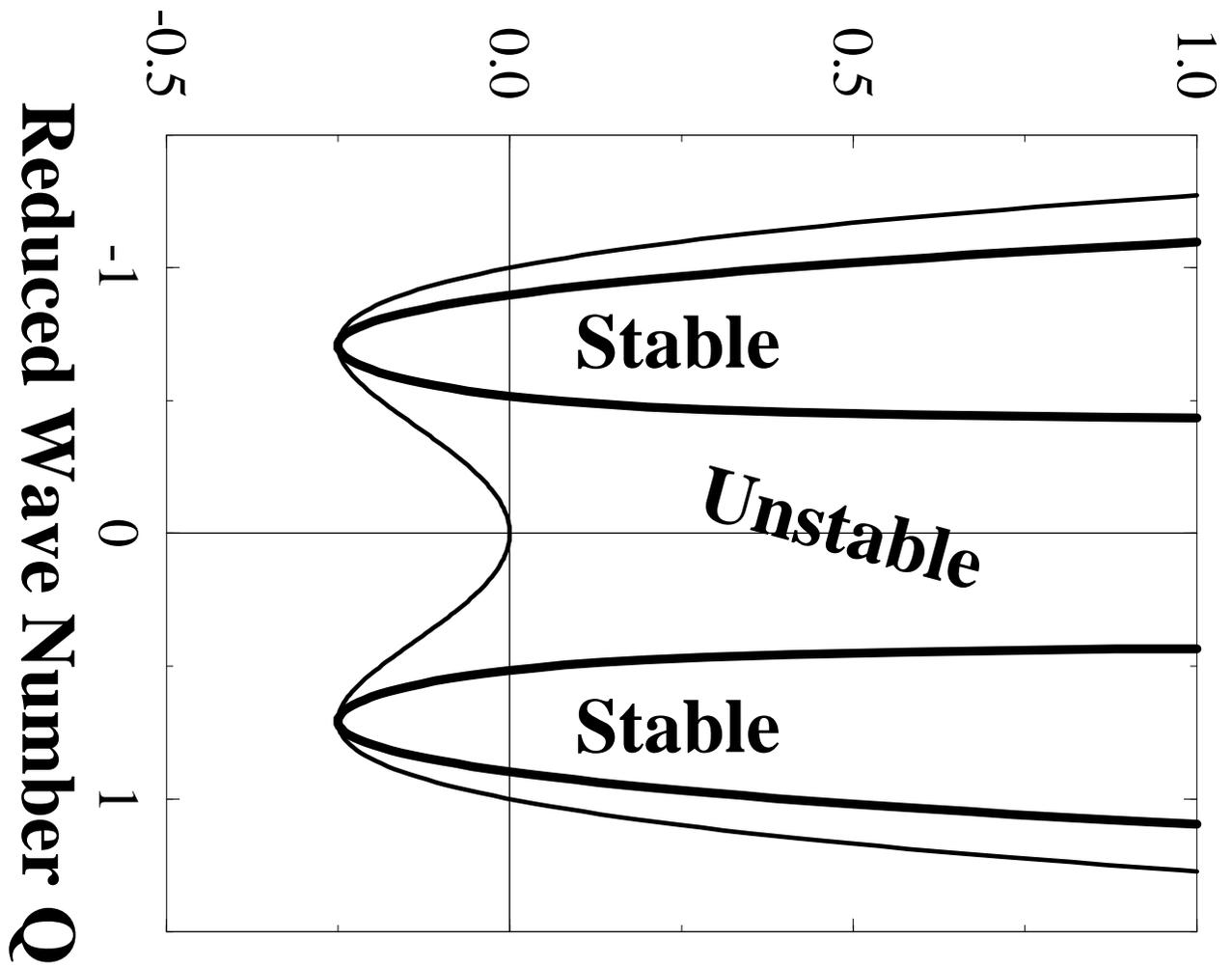
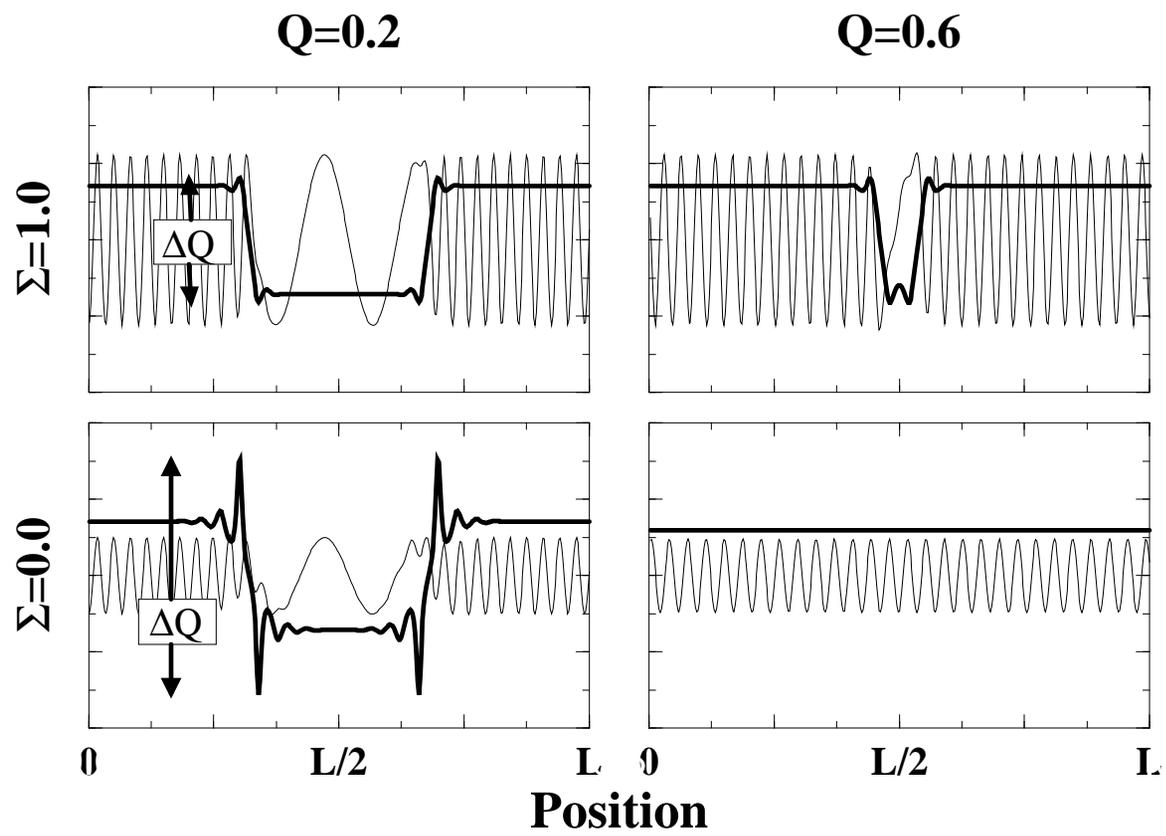
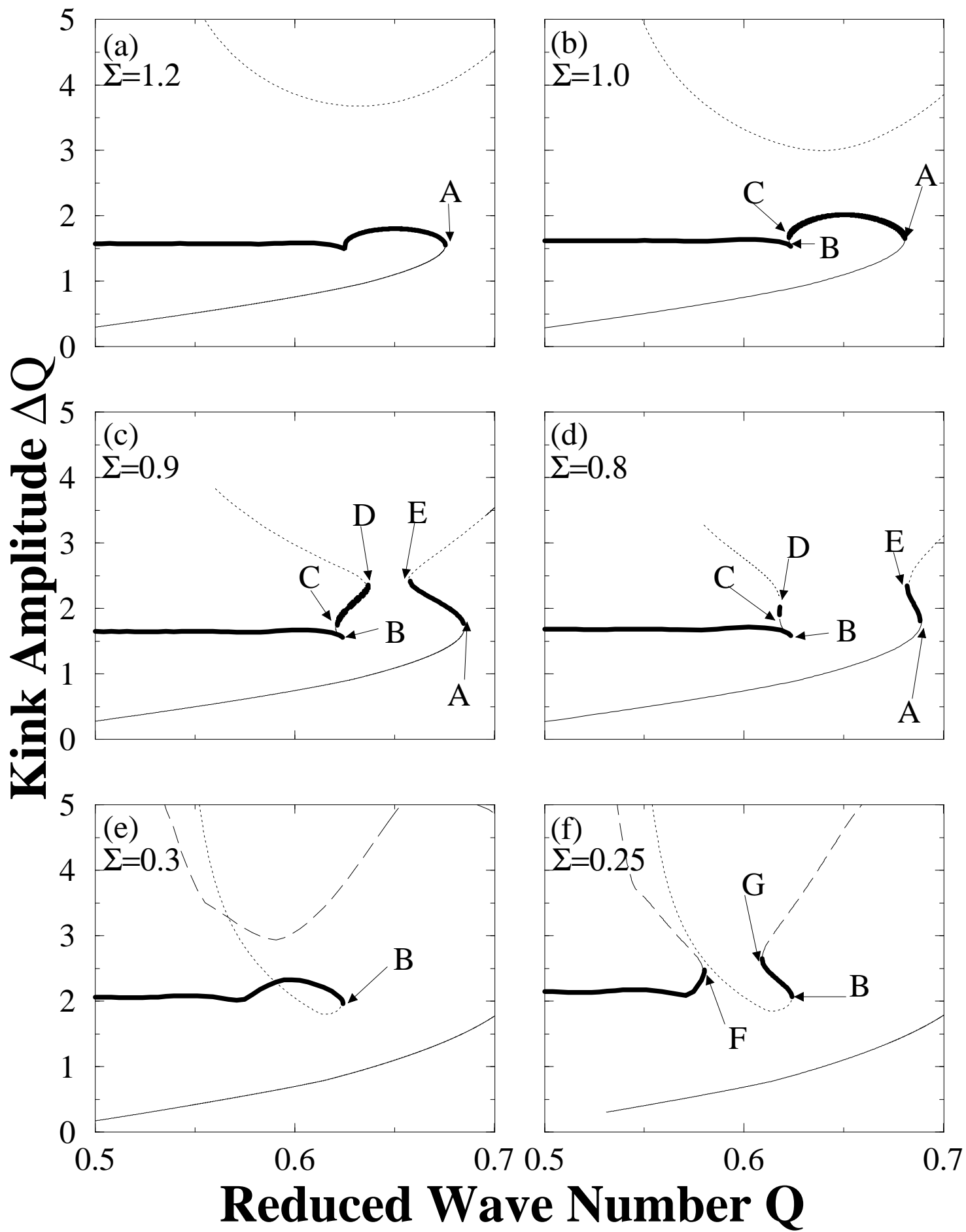
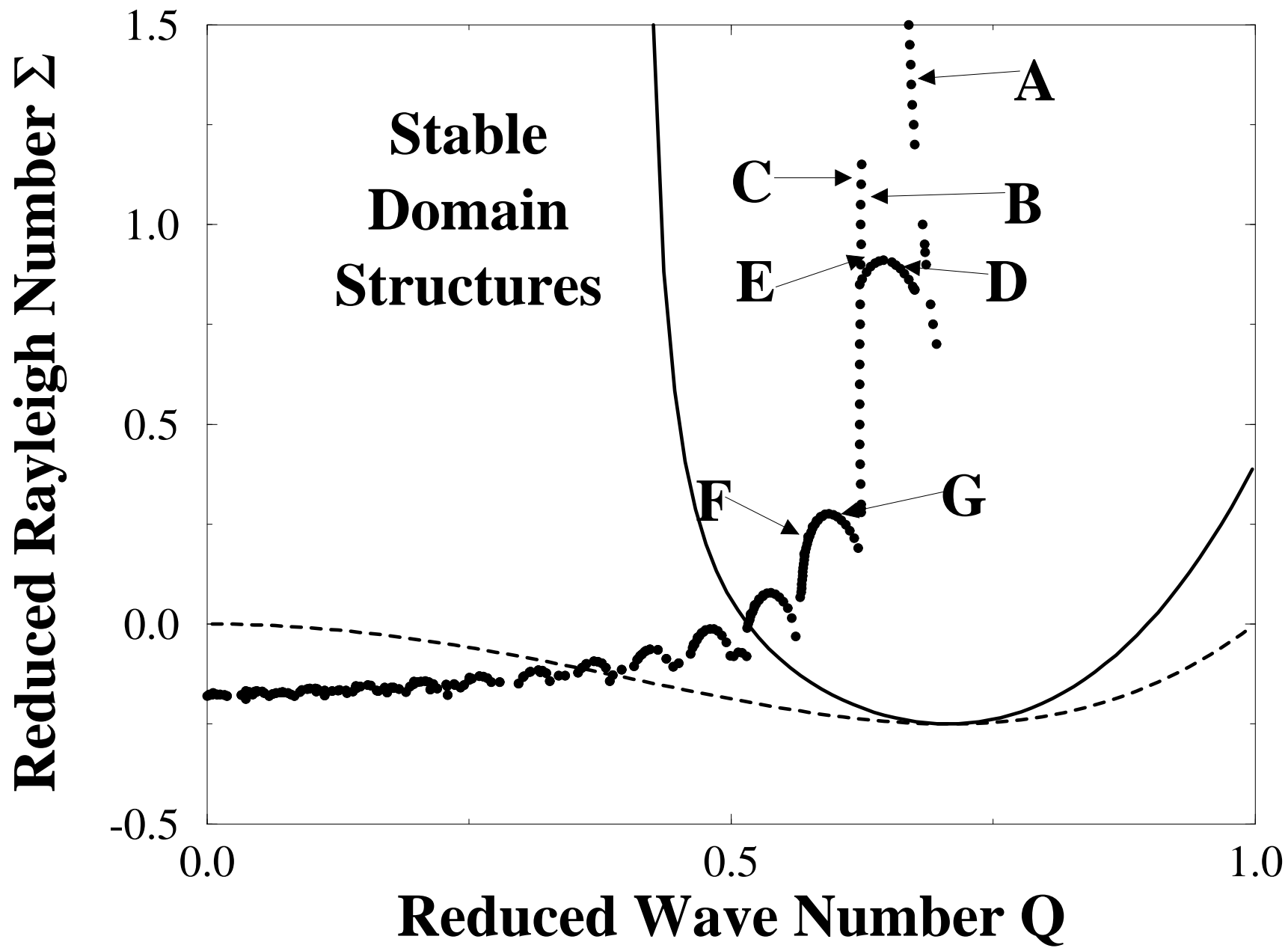
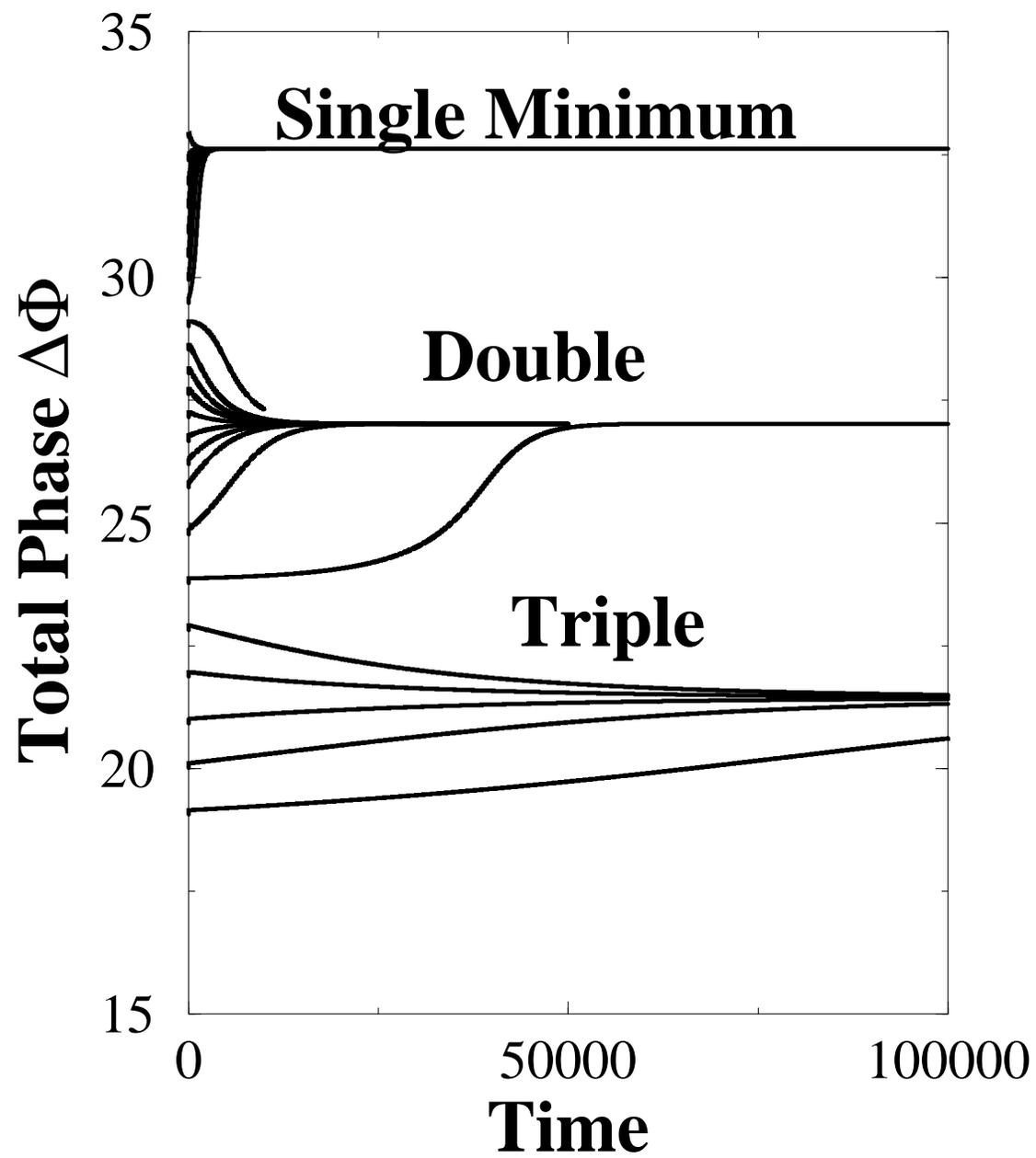
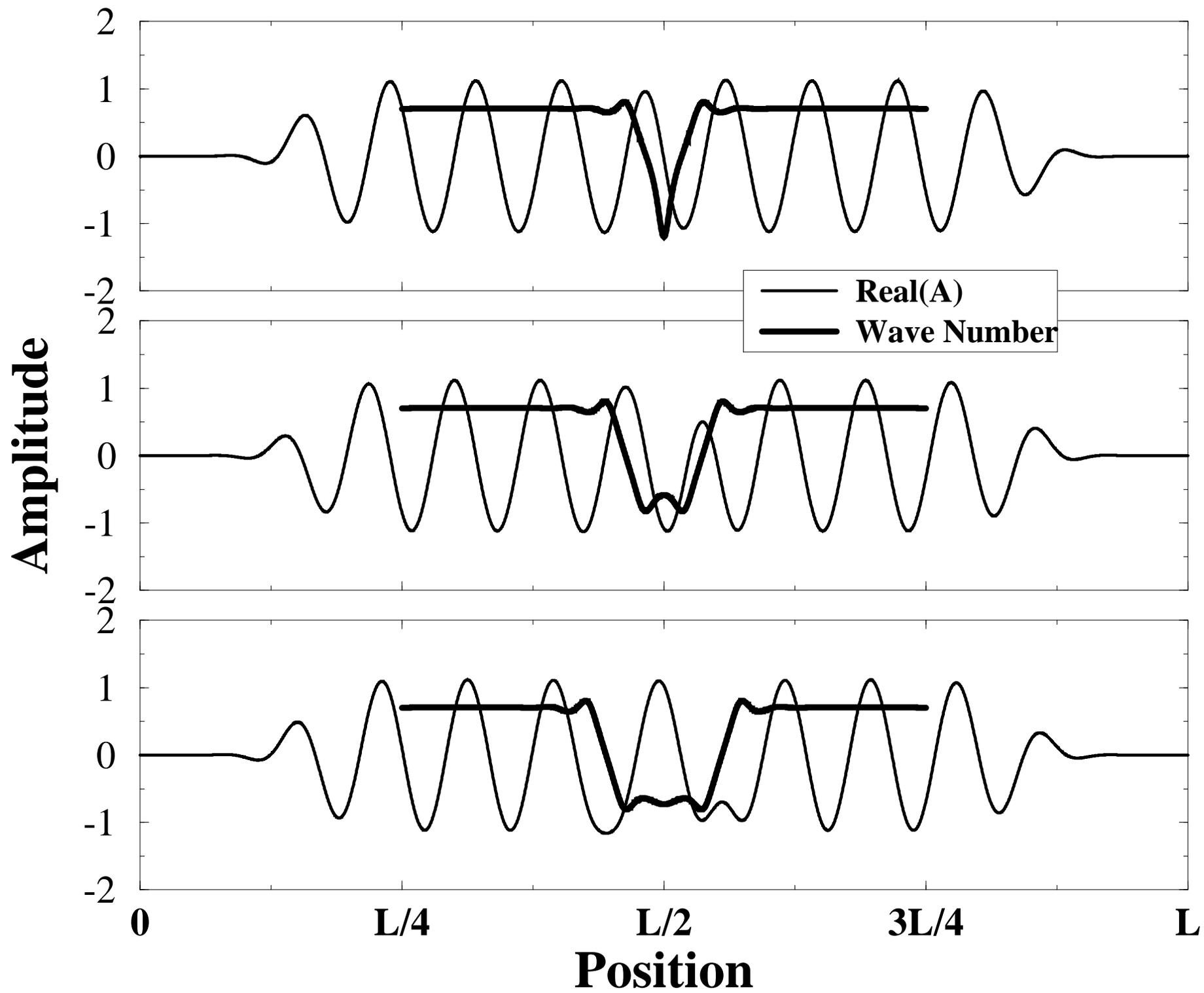
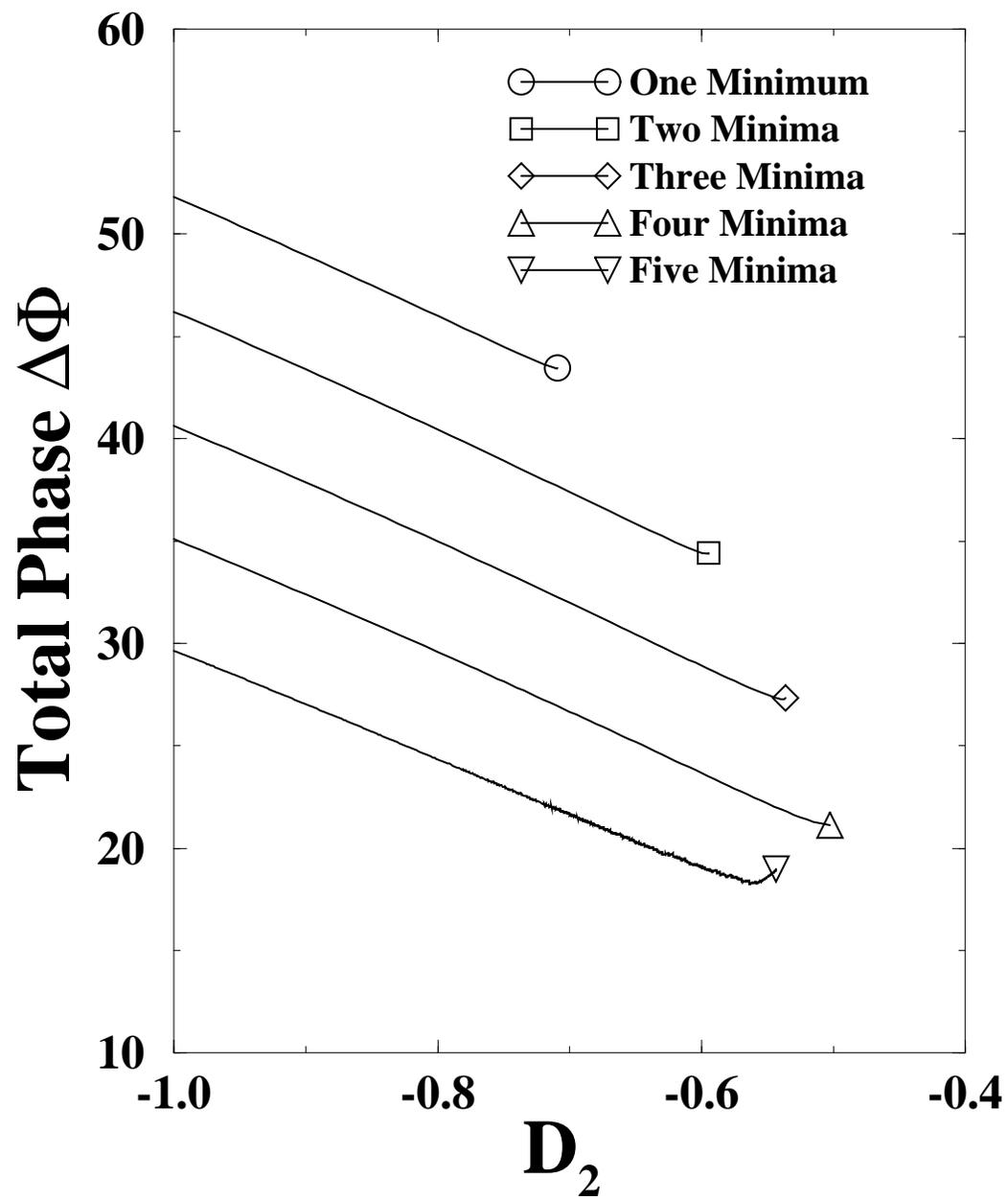
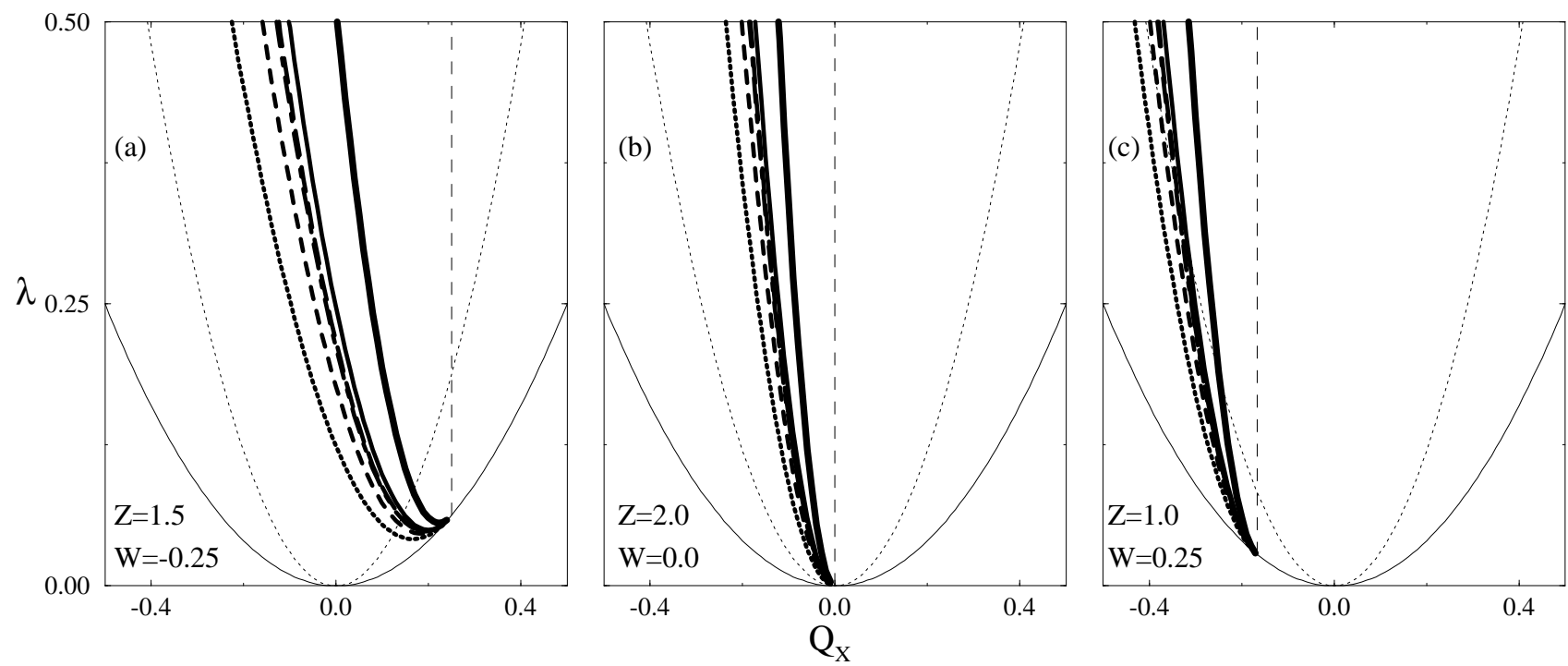
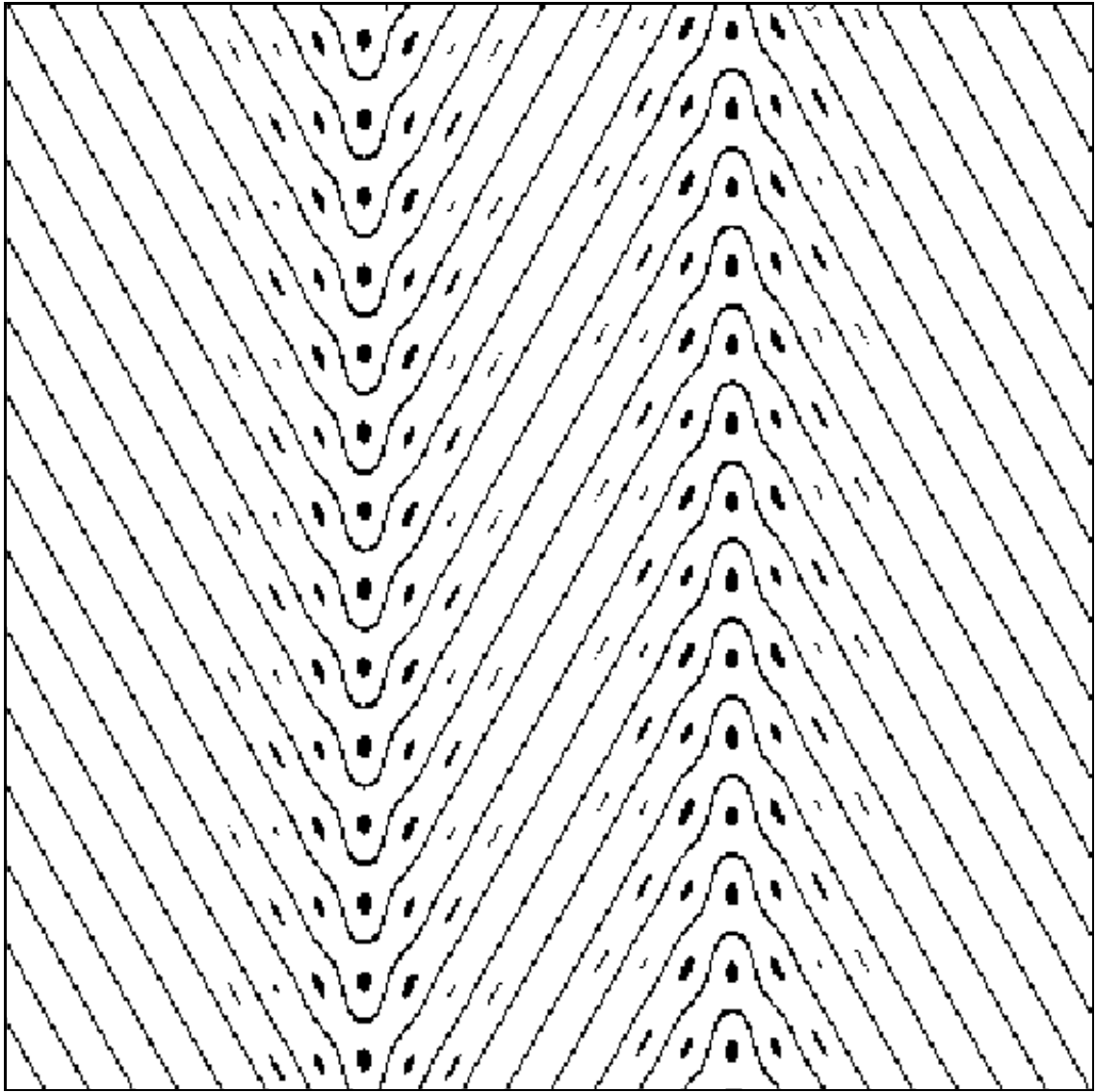
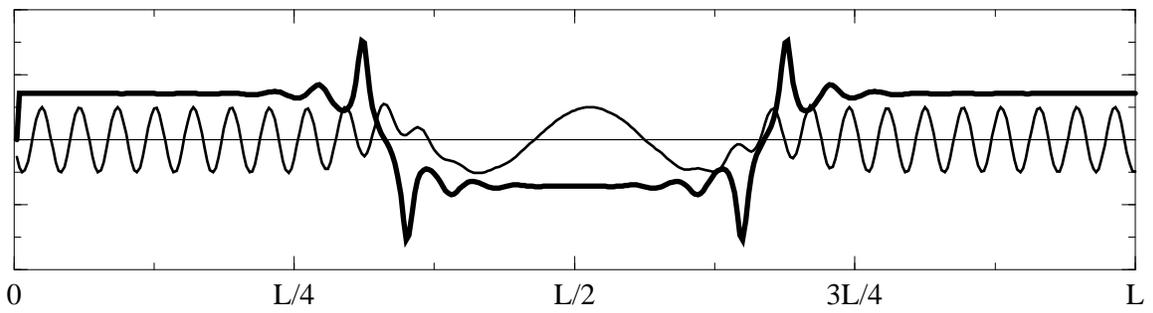
\end{document}